\documentclass[aip,app,graphicx,amsmath,amssymb,reprint]{revtex4-1}
\usepackage{graphicx}% Include figure files
\draft % marks overfull lines with a black rule on the right
\usepackage{dcolumn}% Align table columns on decimal point
\usepackage{bm}% bold math
\usepackage[utf8]{inputenc}
\usepackage[T1]{fontenc}
\usepackage{mathptmx}
\begin{document}

% Use the \preprint command to place your local institutional report number
% on the title page in preprint mode.
% Multiple \preprint commands are allowed.
%\preprint{}

\title{Extreme sensitivity refractive index sensor based on lithography-free metal-dielectric cavity} %Title of paper

% repeat the \author .. \affiliation  etc. as needed
% \email, \thanks, \homepage, \altaffiliation all apply to the current author.
% Explanatory text should go in the []'s,
% actual e-mail address or url should go in the {}'s for \email and \homepage.
% Please use the appropriate macro for the type of information

% \affiliation command applies to all authors since the last \affiliation command.
% The \affiliation command should follow the other information.

\author{Ruoqin Yan}
\author{Tao Wang}
\email{wangtao@hust.edu.cn}%
\author{Xiaoyun Jiang}
\author{Qingfang Zhong}
\author{Xing Huang}
\affiliation{Wuhan National Laboratory for Optoelectronics, Huazhong University of Science and Technology, Wuhan 430074, People’s Republic of China}

%\email[]{Your e-mail address}
%\homepage[]{Your web page}
%\thanks{}
%\altaffiliation{}

% Collaboration name, if desired (requires use of superscriptaddress option in \documentclass).
% \noaffiliation is required (may also be used with the \author command).
%\collaboration{}
%\noaffiliation

\date{\today}

\begin{abstract}
The use of relatively simple structures to achieve high performance refractive index sensors has always been urgently needed. In this work, we propose a lithography-free sensing platform  based on metal-dielectric cavity, the sensitivity of our device can reach 1456700 nm/RIU for solution and 1596700 nm/RIU for solid material, and the FOM can be up to 1234500 /RIU for solution and 1900800 /RIU for solid material, which both are much higher than most sensing methods. This sensor has excellent sensing performance in both TE and TM light, and suitable for integrated microfluidic channels. Our scheme uses a  multi-layers structure with a 10 nm gold film sandwiched between prism and analyte, and shows a great potential for low-cost sensing with high performance.
\end{abstract}

% insert suggested PACS numbers in braces on next line

\maketitle %\maketitle must follow title, authors, abstract and \pacs

% Body of paper goes here. Use proper sectioning commands.
% References should be done using the \cite, \ref, and \label commands

\begin{figure}[ht]
\centering
\includegraphics[scale=0.25]{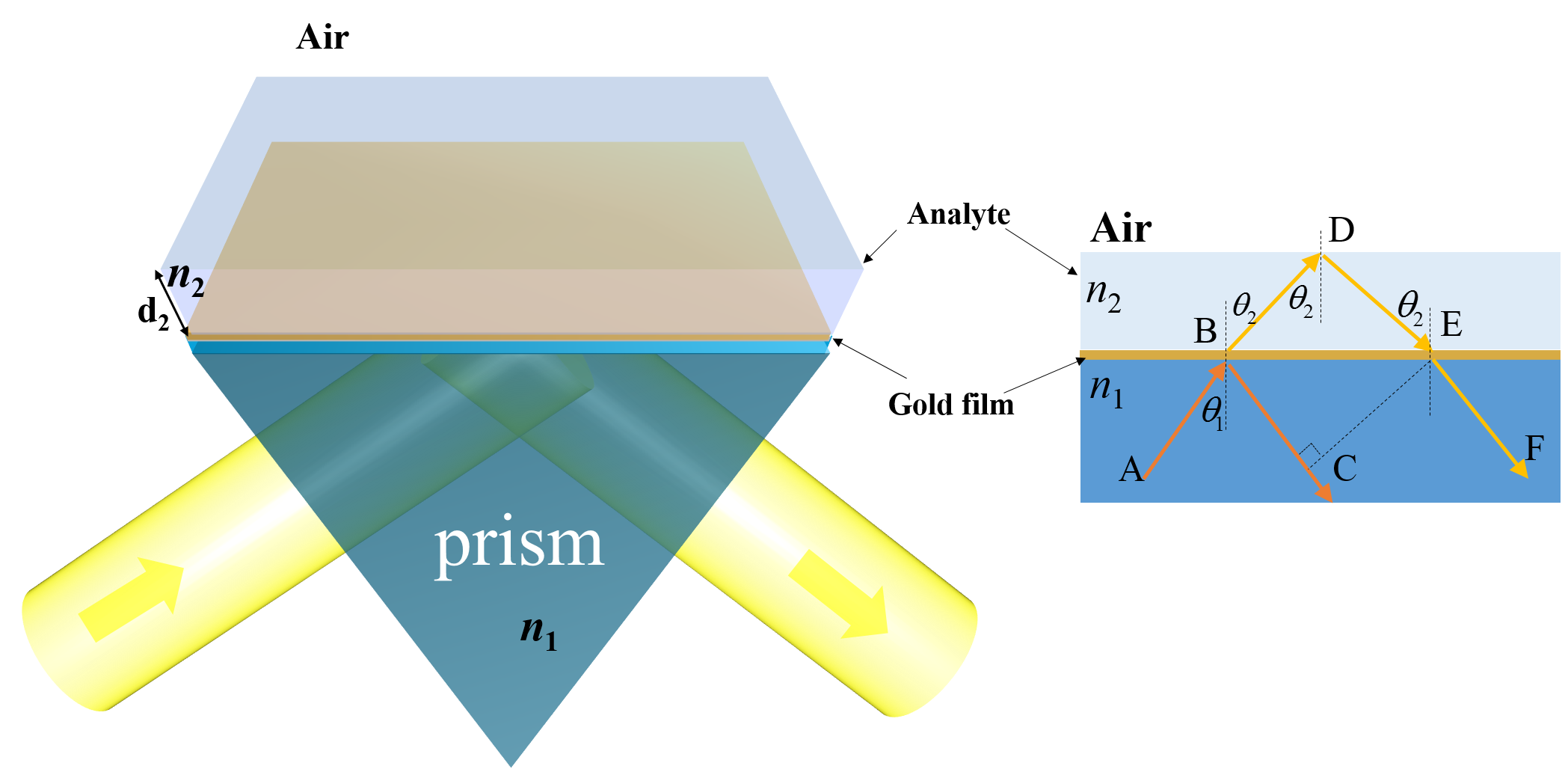}
\caption{\label{fig:model} Schematic drawing of the sensing system.}
\end{figure}

In recent years, optical refractive index sensing methods have attracted plenty of attention for the real-time monitoring of biomolecular binding events because they avoid the time-consuming labeling steps \cite{anker2008biosensing,zanchetta2017label-free}. A variety of refractometric sensing devices have been demonstrated using nanoparticles \cite{mcfarland2003nanoparticle_Ag,chen2008nanoparticle_gold,doherty2017nanoparticle}, metamaterials \cite{wu2012fano,ren2013metamaterial,kabashin2009HMM,sreekanth2016HMM}, photonic crystals \cite{konopsky2007PC,guo2010PC,urbonas2016PC,zhang2016PC}, and narrow-band absorbers \cite{meng2014UNBA,luo2016PNBA,li2017all-metal-grating,chen2018MIM,feng2018mategrating}. It is well known that sensitivity (wavelength shift per refractive index unit, $S=\Delta\lambda/\Delta n$) and figure of merit (FOM) are the two most critical paramters of high-performance sensing devices. The FOM  is defined as ($S/\Delta\omega$), where $\Delta\omega$ is the full-width of the resonant dip at half-maximum, which takes into account the sharpness of the resonance and thus examines the ability to sensitively measure very tiny variation of the resonance wavelength \cite{kabashin2009HMM}. To date, among numerous refractive index sensors have obtained significant progress. For example, Kabashin et al. has used plasmonic nanorod metamaterials for a high sensitivity over 30000 nm/RIU and a FOM up to 330 /RIU at the wavelength of 1.28 $\mu$m \cite{kabashin2009HMM}. Sreekanth et al. has reported a  grating-coupling hyperbolic metamaterials plasmonic sensing platform, which has the sensitivity and FOM up to 30000 nm/RIU and 590 /RIU around the wavelength of 1.3 $\mu$m \cite{sreekanth2016HMM}, respectively. Although the use of metamaterial concept has efficiently improved the sensitivity and FOM of plasmonic sensors, however, these structures rely on intense nanopatterning techniques, which are not cost-effective and hard to manufacture on a large scale. In addition, Feng et al. has proposed an ultranarrow-band absorbers based on asymmetric metagrating structure, which has the sensitivity 1440 nm/RIU and FOM up to 5142.86 /RIU around the wavelength of 1.55 $\mu$m \cite{feng2018mategrating}. Chen et al. has reported a square-patch-based metal–insulator–metal (MIM) structure, which has the sensitivity 1470 nm/RIU and FOM up to 6400 /RIU around the wavelength of 1.48 $\mu$m \cite{chen2018MIM}. These sensors based on narrow-band absorption significantly improved FOM by greatly reducing the full width half maximum (FWHM) of resonance peak, however, their sensitivity is relatively low. Therefore, it is imperative to design high-performance sensors with a lithography-free structure and a lower cost.\\
\indent
In this letter, a simple flat photonic cavity structure for high-performance optical sensing has been proposed, in which thin gold film is sandwiched between a lossless dielectric and a analyte to be measured, as shown in Fig. \ref{fig:model}. By utilizing optical total reflection and high reflectivity of the gold film, we can get the resonance mode with ultra-narrow FWHM. By a series of systematic design and optimization, we have achieved multiple extremely sensitive cavity modes with a maximum sensitivity of  1596700 nm/RIU around the wavelength of 1.42 $\mu$m and a FOM up to 1900800 /RIU, which indicate that the proposed optical sensor is more advanced and promising than conventional plasmonic sensors due to its lithography-free configuration, ultrasensitive performance and extremely high FOM.\\

\indent

The proposed configuration consists of a lossless dielectric prism, a analyte layer to be measured, and a thin gold film layer sandwiched between them. The analyte is exposed to the air. The refractive index of the air, prism, and analyte are $n_{air}$, $n_{1}$, and $n_{2}$, respectively. The thickness of analyte is $d_{2}$. Consider a ray of light from a source A incident on the surface of the gold film at B. Most part of this will be reflected as ray BC and small part refracted in the direction BD. Upon arrival at D, due to total reflection occurs, all of the latter will be reflected to E.  The thin gold film below the analyte has relatively high reflectivity, therefore, the boundary between the analyte and air and the gold film form two mirrors with high reflectivity. The light wave reflected between the two mirrors is constructively or destructively interfered in the analyte, resulting in a series of stationary or standing electromagnetic waves in the analyte. Therefore, the entire structure constitutes a photonic cavity and forms an optical resonator. When the incident light has a wavelength corresponding to one of the cavity modes \cite{kasap2001Optoelectronics},
\begin{equation}\label{eq1}
  q\left( \frac{\lambda }{2{{n}_{2}}} \right)={{d}_{2}}\cos {{\theta }_{2}}\text{=}{{d}_{2}}\sqrt{n_{2}^{2}-n_{1}^{2}{{\sin }^{2}}{{\theta }_{1}}},
\end{equation}
it can oscillate continuously in the resonator and generate transmitted beams. Where $q$ is defined as the cavity mode of the resonator and $\lambda$ is the resonance wavelength. When the refractive index $n_{2}$ changes, the matching condition changes and the resonance wavelength also shifts, which is the basic sensing principle of the proposed device. Define the reflectance of light at the analyte-air interface and analyte-gold film interface, $R_{1}$ and $R_{2}$, respectively. We can calculate the fineness of the cavity \cite{kasap2001Optoelectronics},
\begin{equation}\label{F}
F=\frac{\pi {{R}^{1/2}}}{1-R},
\end{equation}
in which $R$ represents the average reflectance, $R=\sqrt{{{R}_{1}}{{R}_{2}}}$. FWHM =$\lambda_{m}/F$. $F$ increases as $R$ increases, and large finesses lead to sharper mode peaks.\\
\indent
We next investigate the parameter conditions for $\theta _{1}$, $n_{1}$, $n_{2}$, and $d_{2}$ to obtain a considerable value of optical sensitivity. According to Eq.  (\ref{eq1}), we can analyze the wavelength shift $\Delta \lambda$ of resonance mode induced by $\Delta n$ (the refractive index change of the analyte) for the fixed values of $\theta _{1}$ and $d_{2}$ and $n_{1}$, and furthermore, we can deduce the sensitivity formula as below,
\begin{equation}\label{S}
S=\frac{\Delta \lambda }{\Delta n}=\frac{2{{d}_{2}}\left[ \sqrt{{{({{n}_{2}}+\Delta n)}^{2}}-n_{1}^{2}{{\sin }^{2}}{{\theta }_{1}}}-\sqrt{n_{2}^{2}-n_{1}^{2}{{\sin }^{2}}{{\theta }_{1}}} \right]}{q\Delta n}.
\end{equation}
We use K9 glass ($n_{1}$ = 1.5163) as a prism, onto the top of a K9 prism with 10 nm thick gold film. In the simulation, the permittivity of Au is described by a Drude model,
\begin{equation}\label{em}
  \varepsilon_{m}=1-\left[\frac{\omega_{p}^{2}}{\omega(\omega-i/\tau)}\right] ,
\end{equation}
where $\omega_{p}=2\pi \times 2.175\times 10^{15} $ rad/s  is the plasma frequency of Au and $\tau$ is the relaxation time, $1/\tau=2\pi \times 6.5\times 10^{12} $ rad/s .\\
\begin{figure}[ht]
\centering
\includegraphics[scale=0.3]{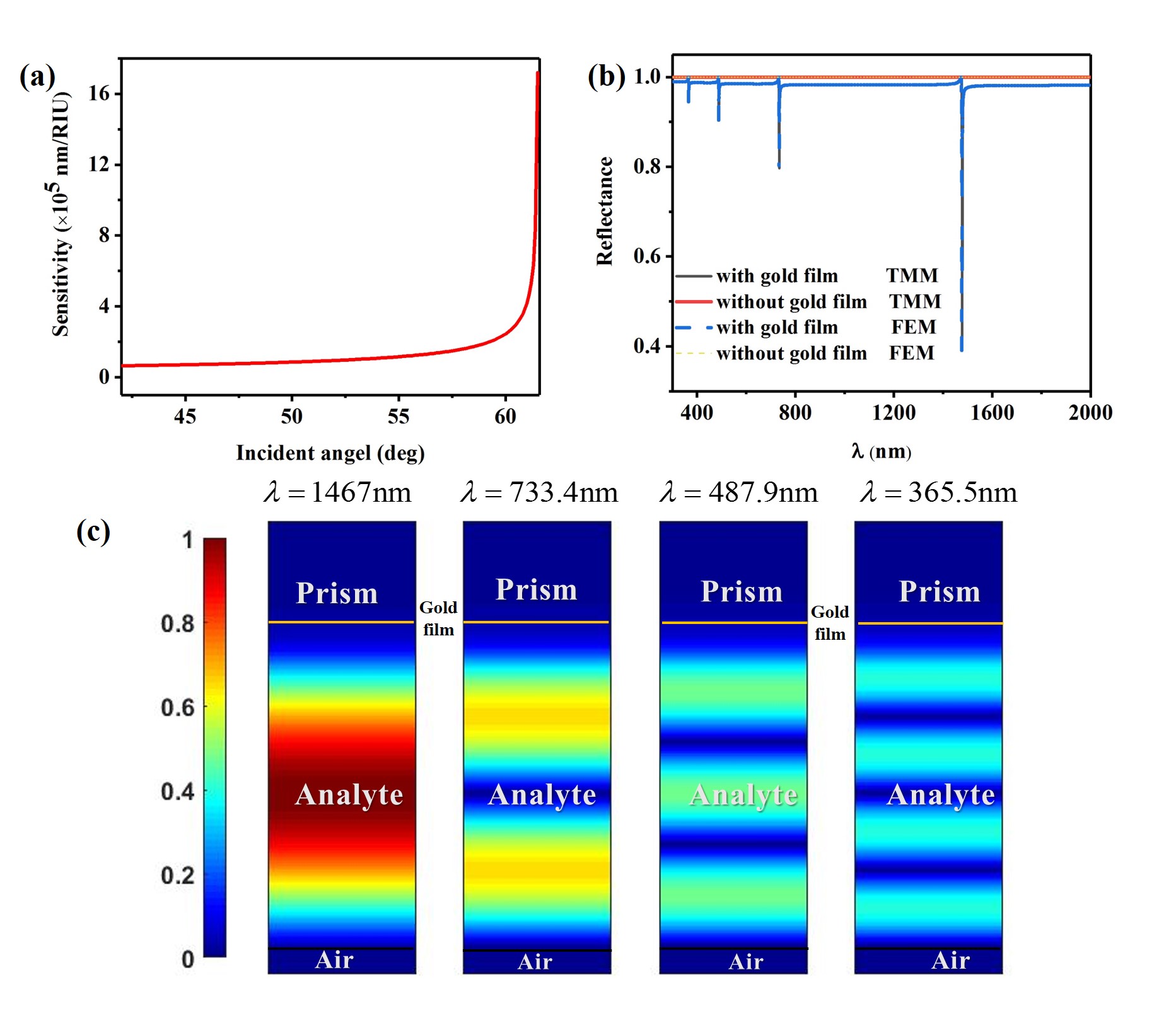}
\caption{\label{fig:figure2a}(a) Sensitivity as functions of $\theta _{1}$, where $n_{1}$, $n_{2}$ and $\Delta n$ are 1.5163, 1.3330, 0.0001 respectively. \label{fig:figure2b}(b) The reflectance comparison between the structure with 10 nm gold film and without 10 nm gold film, which calculated by the transfer matrix method (TMM) and the finite element method (FEM), respectively. \label{fig:figure2c}(c) Normalized electric field distributions corresponding to four resonance wavelengths.}
\end{figure}
\indent
We first analyze the effect of incident angle on sensing performance. In order to ensure the total reflection of the beam between the analyte and the air, the incident angle should satisfy $\theta_{1} >\theta_{C}=\arctan(1/1.5163)=41.27^{\circ}$, and the value of $\theta_{C}$ is only determined
by the refractive index difference between air and prism. As shown in Fig. \ref{fig:figure2a}(a), we fix $n_{1}$= 1.5163, $n_{2}$ = 1.3330, $d_{2}$=21 $\mu$m and $\Delta n$ = 0.0001, and investigate the dependence of the sensitivity on the incident angle. We can find the sensitivity will increase with $\theta_{1}$ before total reflection occurs between prism and gold film. This total reflection angle $\theta^{'}_{C}=\arctan(1.3330/1.5163)=61.536^{\circ}$. When the incident angle is close to $\theta^{'}_{C}$, the sensor can achieve maximum sensitivity. Further, the reflectance comparison between the structure with 10 nm gold film and without 10 nm gold film is plotted in Fig. \ref{fig:figure2b}(b). The results have been achieved by using the transfer matrix method (TMM) and  the finite element method (FEM) \cite{macleod2010TMM,jin2015FEM}. It is clear that TMM results have close agreement with FEM simulation results. As observed in Fig. \ref{fig:figure2c}(b), if the gold film is absent, the whole system will be a transparent. All incident light will be totally reflected at the bottom interface when $\theta_{1}$ > $\theta_{C}$ and the narrow-band resonance peak will disappear (red dash line). Therefore, thin gold film as a mirror with high reflectivity is an indispensable material for high performance optical sensing. The normalized electric field distributions of the resonance mode in our sensor are given in Fig. \ref{fig:figure2c}(c). The wavelengths of the four modes are almost identical to those calculated by Eq.  (\ref{eq1}). This proves the correctness of our theoretical analysis. The light and dark stripes appearing in the incident region are caused by interference. In the analyte region, the light wave strictly satisfies the mode of the cavity, forming a stable standing wave. As shown, no electric field is present in the air.\\
\begin{figure}[ht]
\centering
\includegraphics[scale=0.3]{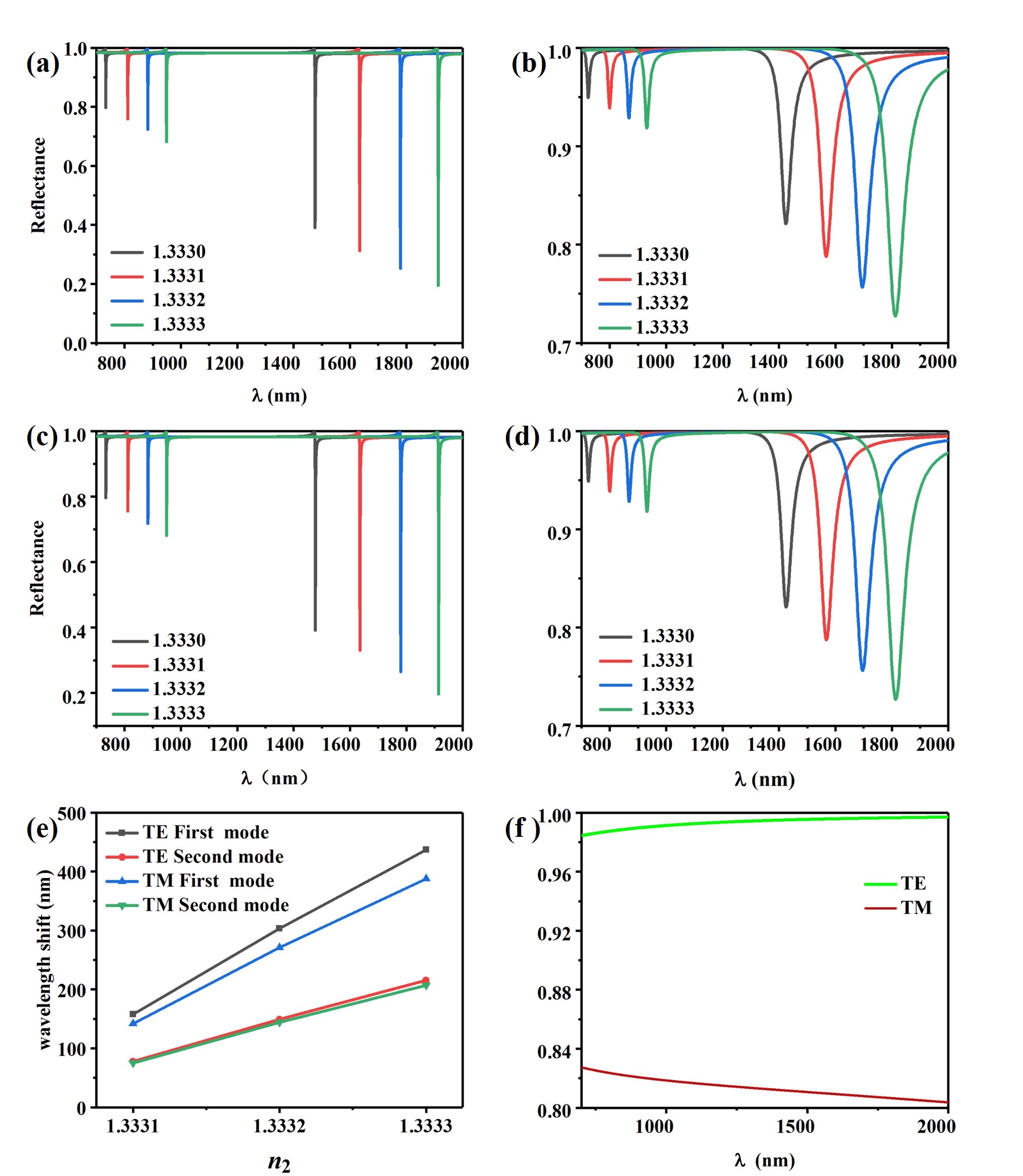}
\caption{ Sensing performance of the proposed structure, (a) and (b) is calculated by TMM and (c) and (d) is calculated by FEM simulation at the  TE and TM incident light, respectively, where $n_{1}$ = 1.5163, $d_{2}$ = 21 $\mu$m, and $\theta_{1}  = 61.5^{\circ}$. (e) The variation of wavelength shift for two modes with different refractive index of analyte. (f) Reflectance of the gold film to TE light and TM light in the analyte.}
\label{fig:figureTETM}
\end{figure}
\indent
We further simulated the sensitivity of TE  and TM  incident light at $61.5^{\circ}$  by TMM and FEM and plotted in Fig. \ref{fig:figureTETM}, respectively, the result of two methods are consistent. As shown in Fig. \ref{fig:figureTETM}(a), by gradually increasing $n_{2}$ from 1.3330 to 1.3333, the resonance mode of TE light for $q = 1$ shows a remarkable red shift from 1476 nm to 1913 nm. Therefore, the deduced sensitivity $S$ is approximately equal to 1456700 nm/RIU, expectedly, due to the frequency selection of the resonator cavity, the FWHM of the resonance peak is 1.18 nm, so the FOM can be as high as 1234500 /RIU , which is an exceptional high value compared with many other sensing methods. For TM light, the resonance wavelength for $q = 1$ drifts from 1424 nm to 1812 nm and the FWHM is 42 nm, so the sensitivity is approximately 1293300 nm/RIU and the FOM is 30793 /RIU. The sensitivity for different resonance modes both TE and TM light are calculated and plotted in Fig. \ref{fig:figureTETM}(e). The sensitivity of the first resonant mode is close to twice the sensitivity of the second resonant mode. This result is in line with the theoretical prediction by Eq. (\ref{S}). We can clearly see that the resonance peak of TM light has a larger FWHM than the TE light. This is because the 10 nm thick gold film has a lower reflectance to the TM light than the TE light, that is $R_{2,TM} < R_{2,TE}$. We can confirm by the following methods. According to the Snell's law of refraction,
\begin{equation}\label{Snell}
\theta_{2}=\arcsin (\frac{{{n}_{1}}}{{{n}_{2}}}\sin{{\theta }_{1}}),
\end{equation}
in which $n_{1}=1.5163$, $n_{2}=1.3330$ and $\theta_{1}=61.5^{\circ}$. So the light reflects back and forth in the cavity at an angle of incidence of $\theta_{2}=88.5^{\circ}$. This angle is greater than the critical angle of total reflection at the analyte and air interface, it can guarantee $R_{1,TM} = R_{1,TE} = 100\% $.
By  simulating the reflection of  the ray DE on the 10 nm thick gold film, we can calculate $R_{2}$ of TE and TM light respectively. As shown in Fig. \ref{fig:figureTETM}(f), $R_{2,TM}$ is less than $ R_{2,TE}$ over the entire waveband. According to Eq. (\ref{F}),  TE light has greater fineness and thus smaller FWHM. This also explains the reason for the high FOM of our sensor,  the total reflection between the analyte and the air to ensure 100\% reflection of light, moreover, the reflectivity of the gold film is high enough. Therefore, the two high-reflectivity mirrors and analyte form a high-quality resonator, only special light that satisfies the phase condition can exist in the cavity. Because the reflectivity of the gold film is not 100\%, part of the light transmitted from the gold film, so we can observe narrow-band formants in the reflection spectrum.\\
 \begin{figure}[ht]
\centering
\includegraphics[scale=0.3]{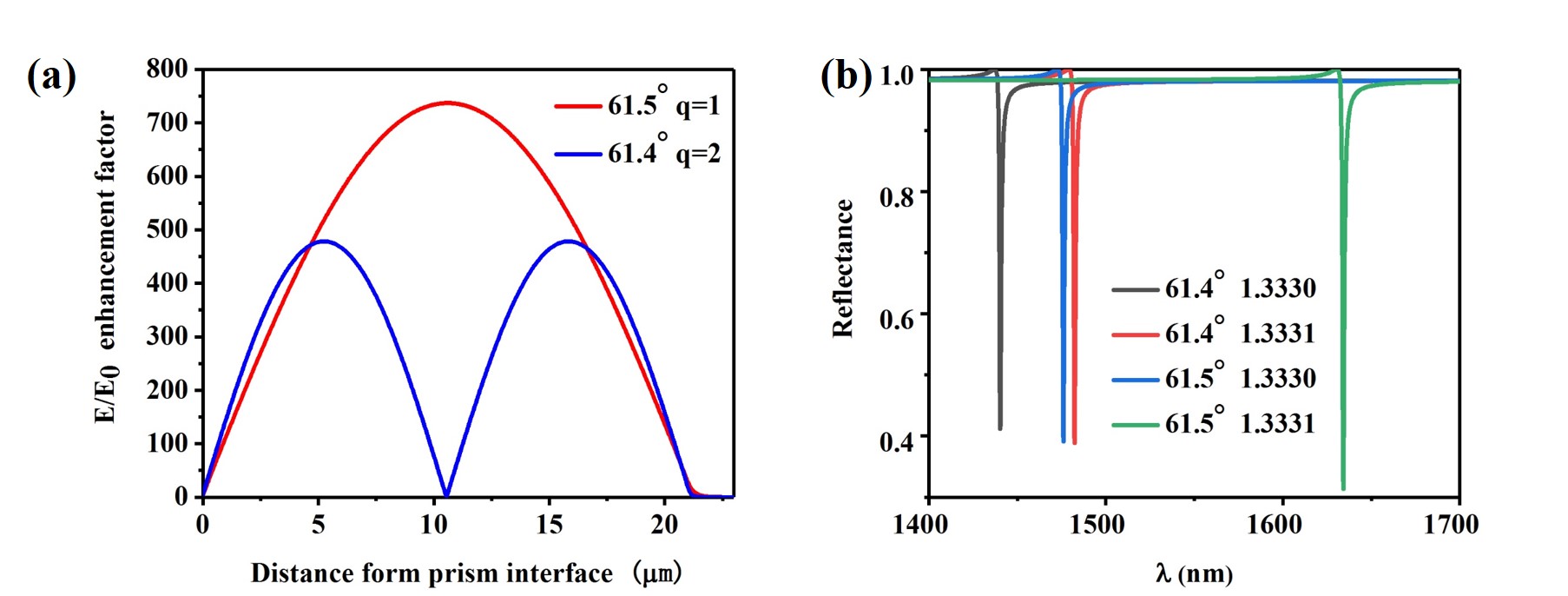}
\caption{\label{fig:figure614}(a) The density distribution of  the electric field  corresponding to resonance wavelength near 1.5 $\mu$m, $\lambda=1476$ nm and $\lambda=1440$ nm are incident at $61.5^{\circ}$ and $61.4^{\circ}$, respectively. $E_{0}$ is  the incident electric field. (b) Sensing performance at $61.5^{\circ}$ and $61.4^{\circ}$ incident angles.}
\end{figure}
\indent
The FOM benefits from higher reflectivity, while high sensing sensitivity results from a very strong electric field in the region of the analyte. It can be seen from Fig. ~\ref{fig:figure2c}(c) that the electric field is the largest at $q=1$ and weakens when $q$ increases. The electric field of the mode of $q=1$ is nearly twice that of the mode of $q=2$, and the sensitivity is also. When the incident angle is close to the total reflection angle $\theta^{'}_{C}$, the electric field in the cavity is maximized, thereby high sensitivity  is obtained.  Fig. \ref{fig:figure614}(a) compares the enhancement factor of  electric field when the resonance mode at $61.5^{\circ}$ and $61.4^{\circ}$ incident angles. The maximum electric field value of $61.5^{\circ}$ is 1.9 times that of $61.4^{\circ}$. As shown in Fig. \ref{fig:figure614}(b), when the refractive index of the analyte changes from 1.3330 to 1.3331, the resonance mode of $61.5^{\circ}$ is red-shifted by 158 nm, and only 42 nm for $61.4^{\circ}$.\\
 \begin{figure}[ht]
\centering
\includegraphics[scale=0.3]{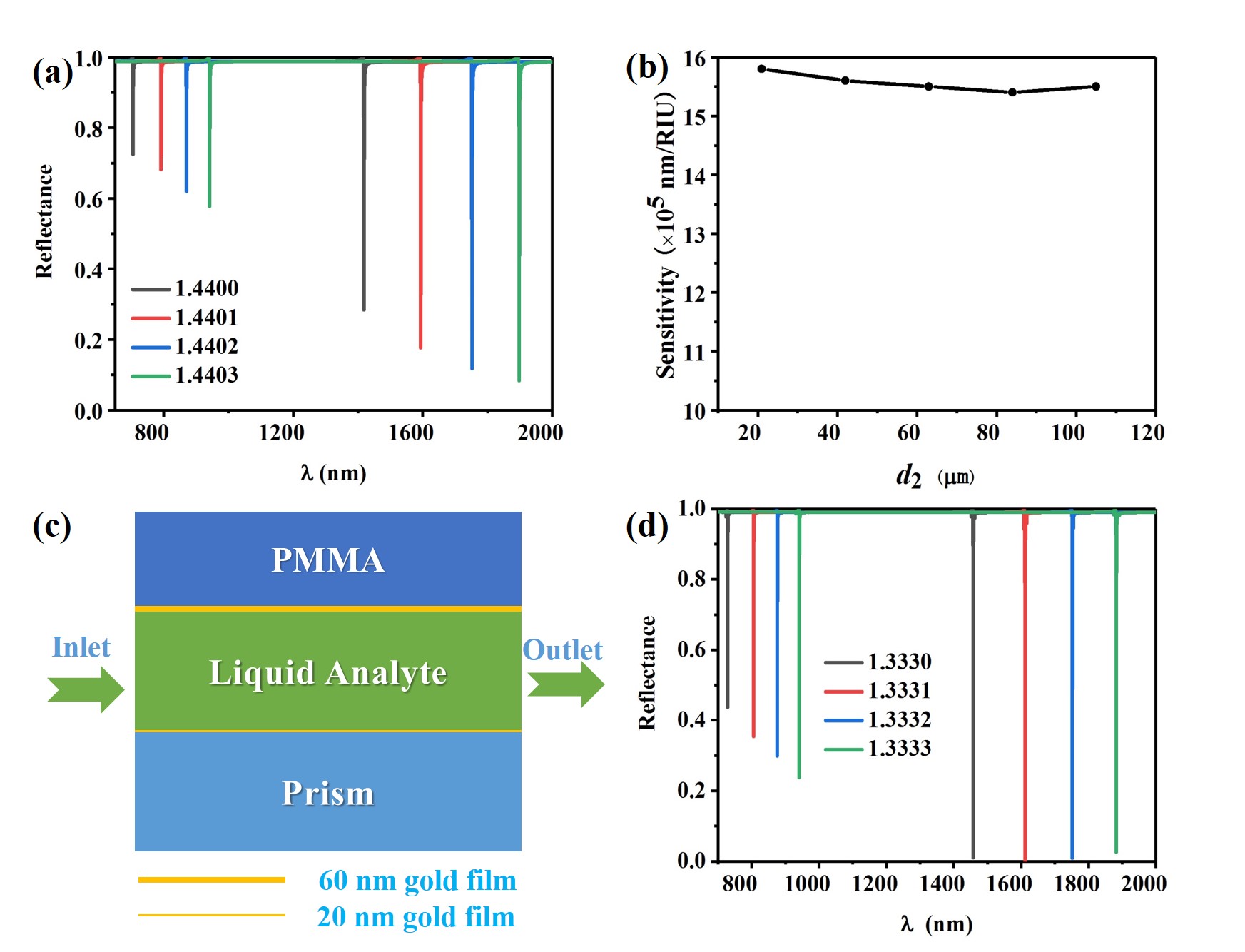}
\caption{\label{fig:figureapply}(a) Refractive index sensing performance for solid materials, where $n_{1}$ = 1.5163, $d_{2}$ = 21 $\mu$m, and $\theta_{1}$ = $71.7^{\circ}$. (b) The sensitivity as functions of $d_{2}$, where $n_{1}$ = 1.5163, $n_{2}$=1.3330, $\Delta n$ =0.0001, and $\theta_{1}$ = $61.5^{\circ}$, and the resonant wavelengths of different mode orders for sensing are fixed around $\lambda$ = 1.5 $\mu$m. (c) A schematic representation of microfluidic channel integrated  nanophotonic cavity. (d) Verify the effect of the microfluidic channel on the sensing performance of the device, where $n_{1}$ = 1.5163, $d_{2}$=21 $\mu$m, and $\theta_{1}$ = $61.5^{\circ}$.}
\end{figure}
\indent
Our sensing devices have excellent operability in practical applications. First of all,  the analytes that can be measured are relatively wide. When the analyte changes from a liquid with a refractive index of 1.3330 to a solid of 1.4400. At this point, just need to adjust the angle of incidence be close to the critical angle of total reflection between the prism and the new analyte. As shown in Fig. \ref{fig:figureapply}(a), by gradually increasing $n_{2}$ from 1.4400 to 1.4403, the resonance mode of TE light for $q = 1$ shows a remarkable red shift from 1418 nm to 1897 nm and the FHWM is 0.84 nm. Therefore, the deduced sensitivity $S$ is approximately equal to 1596700 nm/RIU and the FOM can be as high as 1900800 /RIU. Secondly, the thickness of analyte  can be adjusted according to an actual measurement. When the other conditions are fixed, only the thickness of the analyte layer is changed from 21 $\mu$m to 105 $\mu$m. The sensitivity as functions of $d_{2}$  when  $n_{1}$ changed from 1.3330 to 1.3331 are calculated and plotted in Fig. \ref{fig:figureapply}(b), where the resonance wavelengths of different modes for sensing are fixed around $\lambda$= 1.5 $\mu$m, the sensitivity value slightly changes above 1550000 nm/RIU. Thirdly, man may worry  it is difficult to accurately guarantee the same thickness of liquid analyte  for each measurement. In order to avoid this error, we can use a polymethylmethacrylate (PMMA, n=1.49) microfluidic channel integrated photonic cavity \cite{sreekanth2019microfluidics}. Based on the original structure, we introduced a 2 mm PMMA layer between the analyte and the air. The side of the PMMA layer in contact with the analyte is plated with a 60 nm thick gold film. The sensor structure integrated with the microfluidic channel is shown in Fig. \ref{fig:figureapply}(c). The sensitivity are simulated and plotted in Fig. \ref{fig:figureapply}(d). High sensing performance is not attenuated by microfluidic channels, which provides a good solution for measuring liquid analytes in our practical applications. \\
\indent
Finally, we compare the proposed structure with several representative sensors reported recently in terms of sensing performance. As shown in Tab.\ref{tab:table1}, both of the sensitivity and FOM of our proposed sensing structure have been improved dramatically compared with previous sensing research based on plasmonics, metamaterials, narrow-band absorbers and graphene, moreover, our structure is easier to make, which indicate that our design is extremely competitive for ultrahigh-sensing performance.\\

\newcommand{\tabincell}[2]{\begin{tabular}{@{}#1@{}}#2\end{tabular}}
\begin{table}[ht]
\vspace{-0.5cm}
\centering
\caption{  \label{tab:table1}\bf  Performance Comparisons between  Our Work and Recent Reported Prominent Optical Sensors }
\setlength{\tabcolsep}{3mm}{
\begin{tabular}{cccc}
\hline
methods & \tabincell{c}{central $\lambda$\\  ($\mu$m)}  & \tabincell{c}{sensitivity \\ (nm/RIU)}  & \tabincell{c}{FOM \\ (/RIU)} \\
\hline
plasmonics\cite{otte2009plasmonic} & 1 & 13000 & 138 \\
metamaterial\cite{kabashin2009HMM} & 1.28 & 30000 & 330 \\
metamaterial\cite{sreekanth2016HMM} & 1.3 & 30000 & 590 \\
metagrating\cite{feng2018mategrating} & 1.55 & 1440 & 5143 \\
MIM\cite{chen2018MIM} & 1.48 & 1470 & 6400 \\
graphene\cite{li2019graphene} & 1 & 440000 & 6833 \\
this work (liquid) & 1.47 & 1456700 &  1234500 \\
this work (solid) & 1.42 & 1596700 &  1900800 \\
\hline
\end{tabular}}
\end{table}
\indent
In conclusion, a polarization-insensitive refractive index sensor based on lithography-free planar photonic cavity has been theoretically and numerically demonstrated.  Benefit from the strong electric field and narrow-band resonance mode generated by the photonic cavity configuration, the sensitivity as high as 1456700 nm/RIU for liquid analyte and 1596700 nm/RIU for solid analyte is achieved, and the FOM can reach  1234500 /RIU for  liquid analyte and 1900800 /RIU for solid analyte. Furthermore, the current sensor design is highly scalable including analyte thickness and detection range. Since the developed sensor device is a lithography-free design, the sensing area can be significantly enlarged and fabrication costs can be dramatically reduced. Therefore, our method is promising for developing cost-effective high-performance in biological or chemical sensing.
\begin{acknowledgments}
 Financial support from the National Natural Science Foundation of China (NSFC) (61775064) is gratefully acknowledged.
 \end{acknowledgments}
% If in two-column mode, this environment will change to single-column format so that long equations can be displayed.
% Use only when necessary.
%\begin{widetext}
%$$\mbox{put long equation here}$$
%\end{widetext}

% Figures should be put into the text as floats.
% Use the graphics or graphicx packages (distributed with LaTeX2e).
% See the LaTeX Graphics Companion by Michel Goosens, Sebastian Rahtz, and Frank Mittelbach for examples.
%
% Here is an example of the general form of a figure:
% Fill in the caption in the braces of the \caption{} command.
% Put the label that you will use with \ref{} command in the braces of the \label{} command.
%
% \begin{figure}
% \includegraphics{}%
% \caption{\label{}}%
% \end{figure}

% Tables may be be put in the text as floats.
% Here is an example of the general form of a table:
% Fill in the caption in the braces of the \caption{} command. Put the label
% that you will use with \ref{} command in the braces of the \label{} command.
% Insert the column specifiers (l, r, c, d, etc.) in the empty braces of the
% \begin{tabular}{} command.
%
% \begin{table}
% \caption{\label{} }
% \begin{tabular}{}
% \end{tabular}
% \end{table}

% Create the reference section using BibTeX:
\bibliography{aiptemplate}

\end{document}